\documentclass[showpacs,preprint,floatfix,preprintnumbers,nobibnotes,superscriptaddress]{revtex4}
\usepackage{amsmath,amssymb,amsbsy,color}
\usepackage{epsfig}

\begin{document}
\preprint{KEK-CP-232}
\preprint{OU-HET-654-2010}

\title{
  Strong coupling constant from vacuum polarization functions in
  three-flavor lattice QCD with dynamical overlap fermions
}

\author{E. Shintani}
\email{shintani@het.phys.sci.osaka-u.ac.jp}
\affiliation{Department of Physics, Osaka University, Toyonaka, Osaka 560-0043 Japan}

\author{S. Aoki}
\affiliation{Graduate School of Pure and Applied Sciences, University of Tsukuba, 
Tsukuba, Ibaraki 305-8571, Japan}

\author{H. Fukaya}
\affiliation{Department of Physics, Nagoya University, Nogaya 464-8602, Japan}

\author{S. Hashimoto}
\affiliation{KEK Theory Center, Institute of Particle and Nuclear
             Studies, High Energy Accelerator Research Organization (KEK)}
\affiliation{School of High Energy Accelerator Science, The Graduate
        University for Advanced Studies (Sokendai), Ibaraki 305-8081,Japan}

\author{T. Kaneko}
\affiliation{KEK Theory Center, Institute of Particle and Nuclear
             Studies, High Energy Accelerator Research Organization (KEK)}
\affiliation{School of High Energy Accelerator Science, The Graduate
        University for Advanced Studies (Sokendai), Ibaraki 305-8081,Japan}

\author{T. Onogi}
\affiliation{Department of Physics, Osaka University, Toyonaka, Osaka 560-0043 Japan}

\author{N. Yamada}
\affiliation{KEK Theory Center, Institute of Particle and Nuclear
             Studies, High Energy Accelerator Research Organization (KEK)}
\affiliation{School of High Energy Accelerator Science, The Graduate
        University for Advanced Studies (Sokendai), Ibaraki 305-8081,Japan}

\collaboration{JLQCD collaboration}

\begin{abstract}
  We determine the strong coupling constant $\alpha_s$ from a
  lattice calculation of vacuum polarization functions (VPF)  
  in three-flavor QCD with dynamical overlap fermions. 
  Fitting lattice data of VPF to the continuum perturbative formula 
  including the operator product expansion, we extract the QCD scale
  parameter $\Lambda_{\overline{MS}}^{(3)}$. 
  At the $Z$ boson mass scale, we obtain 
  $\alpha_s^{(5)}(M_Z)=0.1181(3)(^{+14}_{-12})$, where the first
  error is statistical and the second is our estimate of various
  systematic uncertainties.
\end{abstract}
\maketitle

\section{Introduction}

Strong coupling constant $\alpha_s$ is one of the fundamental
parameters in the Standard Model.
Its precise determination from various sources provides a crucial test
of Quantum Chromodynamics (QCD).
Experimentally, it is obtained through high energy particle
scatterings involving quarks, such as
$e^+e^-\to \mbox{hadrons}$ (see \cite{Bethke:2009, Amsler:2008zzb} for a summary),
for which perturbative calculation of QCD is possible.

Among other measurements, the hadronic decay rate of tau lepton
\cite{Davier:2005xq} provides one of the most precise determinations of $\alpha_s$. 
The tau-lepton hadronic decay rate is written in terms of a vacuum
polarization function (VPF) of weak currents.
Since the perturbative QCD calculation cannot be directly applied
for physical time-like momentum transfer (or the virtual $W$ boson mass)
$q^2$, one considers a total decay rate that involves an integral over $q^2$.
The decay rate is then calculated using three-loop perturbative
expansion and the Operator Product Expansion (OPE) to parametrize
corrections at low energies.
The final error in $\alpha_s$ contains those from the
perturbative expansion and uncertainties of condensates appearing in
OPE, both of which are non-negligible because the relevant energy
scale is of order of 1~GeV, {\it i.e.} lower than the tau lepton mass
$m_\tau$.  
There is also an assumption of the quark-hadron duality, which is not
trivially satisfied. 

Theoretically, the perturbative calculation and the extraction of
$\alpha_s$ become more transparent if the {\it experimental} data for
VPF are available at space-like momenta. 
Although there is no such direct measurement, lattice QCD is able to
provide a non-perturbative {\it calculation} of VPF at space-like momenta. 
Since the calculation is based on the first-principles of QCD, for
which the input parameters are low-lying hadron spectrum, it gives
another method to extract $\alpha_s$. 

Using lattice QCD, one can calculate VPF of vector ($V$) and
axial-vector ($A$) channels from two-point functions of
those currents.
After a Fourier transformation in four space-time dimensions, VPF can
be obtained as a function of space-like momentum squared $Q^2$ ranging
from zero to the order of lattice cutoff squared. 
Our recent lattice study with two flavors of dynamical overlap
fermions demonstrated that such data can indeed be used to extract
$\alpha_s$ combining with a perturbative calculation in the continuum
scheme \cite{Shintani:2008ga}. 
The result is consistent with other lattice calculations in
two-flavor QCD \cite{DellaMorte:2004bc,Gockeler:2005rv}.

One of the main advantages of our method is that the calculation can
be done on existing gauge configurations produced for light-hadron
spectrum \cite{Aoki:2008tq}.
Furthermore, unlike other methods previously applied, there is no need
of multi-loop perturbative calculation on the lattice, which is so
complicated that one typically has to develop automated tool dedicated
for a given lattice action.
Finite volume effect is negligible for VPF at relatively large $Q^2$,
and discretization effect is carefully studied and is shown to be
under control. 
The method therefore provides a reliable and economical way to extract
the most important parameter of QCD, {\it i.e.} $\alpha_s$. 
A similar idea has also been applied for the charmonium two-point function
to determine $\alpha_s$ as well as charm quark mass \cite{Allison:2008xk}.

In this work, we extend our previous study to the case of realistic
three-flavor QCD with dynamical light and strange quarks.
We also improve the calculation by employing the conserved current for
the overlap fermion \cite{Kikukawa:1998py}, which simplifies the
possible form of the two-point function as it satisfies the
Ward-Takahashi (WT) identity on the lattice.
The extraction of VPF is thus made more straightforward. 
The determination of the strong coupling constant uses the continuum
perturbative QCD formula up to four loops and OPE up to $1/Q^4$ terms.
The value of the quark condensate is calculated independently and used
as an input in this work.
The result is translated to the common definition, {\it i.e.}
$\alpha_s^{(5)}(M_Z)$, the strong coupling constant of five-flavor
QCD in the $\overline{\mathrm{MS}}$ scheme at the $Z$ boson mass scale.

Numerical simulations of lattice QCD are carried out with 2+1-flavors
of dynamical fermions described by the overlap fermion formulation
\cite{Neuberger:1997fp,Neuberger:1998wv}.
We have data at a single lattice spacing with the inverse lattice 
spacing $a^{-1}$ = 1.83(1)~GeV estimated using the static quark potential
with an input for the Sommer scale $r_0$ = 0.49~fm
(the associated error for this is discussed later).
The lattice volume is $16^3\times 48$, which leads to the physical
volume about (1.8~fm)$^3\times$(5.4~fm).
The gauge configurations are generated in the course of the dynamical
overlap fermion simulations by the JLQCD and TWQCD collaborations
\cite{Hashimoto:2008fc}.
The set of up and down quark masses $m_{ud}$ covers the range of
(0.2--0.8)$m_s^{\mathrm{phys}}$ and the set of the strange quark mass $m_s$
covers the range of (1.0--1.3)$m_s^{\mathrm{phys}}$ with
$m_s^{\mathrm{phys}}$ the physical strange quark mass. 
The valence quark mass in the calculation of VPF is set equal to the
up and down sea quark mass.
For each combination of $m_{ud}$ and $m_s$, we use 260 configurations,
each of which is separated by 100 HMC trajectories.
Global topological charge of the gauge configurations is fixed to zero,
which may induce small finite size effect for long distance physical
quantities \cite{Aoki:2007ka}. 
We expect that this gives negligible effects on the short distance
physics considered in this work.

The rest of the paper is organized as follows.
In Section~\ref{sec:lattice}, we describe the details of the lattice
calculation of VPF, including the definition of the overlap fermion
used in this work and a construction of the conserved current for the
overlap fermion.
Section~\ref{sec:fit} discusses the fit of VPF to the perturbative
formula. 
The possible systematic errors are discussed in
Section~\ref{sec:error}, followed by our final results.
Our conclusions are given in Section~\ref{sec:conclusion}.

\section{Lattice calculation of vacuum polarization functions}
\label{sec:lattice}
In the continuum theory,
transverse ($\Pi_J^{(1)}(Q)$) and longitudinal
($\Pi_J^{(0)}(Q)$) parts of VPF are defined through two-point
functions $\langle J^a_\mu(x) J^b_\nu(0)\rangle$ of either 
vector ($J_\mu=V_\mu$) or axial-vector ($J_\mu=A_\mu$) currents
with $a$, $b$ the flavor indices.
Namely, after a Fourier transformation to the momentum space, the
two-point functions are parametrized as
\begin{equation}
  \langle J^a_\mu J^b_\nu\rangle(Q) = \delta^{ab}
  \left[
    (\delta_{\mu\nu}Q^2-Q_\mu Q_\nu)\Pi_J^{(1)}(Q) 
    - Q_\mu Q_\nu \Pi_J^{(0)}(Q) 
  \right],
\end{equation}
where the momentum $Q_\mu$ is space-like as we work on an Euclidean 
space-time lattice. 
Because of the WT identities, the longitudinal part of the
vector channel vanishes, $\Pi_V^{(0)}(Q)=0$,  
while the axial-vector channel is proportional to the quark mass.

On the lattice, we employ the overlap fermion formulation
\cite{Neuberger:1997fp,Neuberger:1998wv}, whose action
$S_{\rm ov}=\sum_{x,y}\bar q(x)D_{\rm ov}(x,y)q(y)$ 
is specified by the massive overlap-Dirac operator 
\begin{equation}
  \label{eq:overlap}
  D_{\rm ov}(x,y) = \left(m_0+\frac{m}{2}\right) +
  \left(m_0-\frac{m}{2}\right) \gamma_5 \mathrm{sgn}[H_W(x,y;-m_0)]
\end{equation}
for a quark mass $m$.
Here, $m_0$ is a parameter to define the overlap kernel
$H_W(x,y;-m_0)=\gamma_5 (D_W(x,y)-m_0)$ with $D_W(x,y)$ the
conventional Wilson-Dirac operator.
In this study we take $m_0=1.6$. 
(Here and in the following we set $a=1$, unless otherwise stated.)
The overlap action has an exact symmetry under a chiral rotation
defined with the modified chirality operator
$\hat\gamma_5(x,y)\equiv\gamma_5(\delta_{x,y}-D_{\rm ov}(x,y)/{m_0})$,
so that the continuum-like axial WT identities are hold on the lattice
at finite lattice spacings.

The conserved vector current for this action has a complicated form,
which can be written in a general form 
$V^{0,\rm cv}_\mu(x) = \sum_{w,z}\bar q(w)K_\mu(w,z|x)q(z)$
with a non-local kernel $K_\mu(w,z|x)$.
$K_\mu(w,z|x)$ is determined such that it forms a Noether current 
under a local vector transformation 
\begin{eqnarray}
  \delta S_{\rm ov} &=& \sum_{x,y}\bar q(x) \left[
    -\alpha(x)D_{\rm ov}(x,y)+D_{\rm ov}(x,y)\alpha(y)
    \right] q(y)
  \nonumber\\
  &=& \sum_{x,y,z} \bar q(x)\alpha(z)\partial^{z\,*}_\mu K_\mu(x,y|z)q(y)
\end{eqnarray}
with $\alpha(x)$ a local parameter \cite{Kikukawa:1998py}.
The derivative $\partial_\mu^{x\,*}$ denotes
a backward derivative operator
$\partial_\mu^{x\,*} V_\mu(x) \equiv V_\mu(x)-V_\mu(x-a\hat\mu)$
on the lattice.
Similarly, for flavor non-singlet transformations, we can derive
flavor non-singlet conserved vector and axial-vector currents as
\begin{eqnarray}
  V_\mu^{a,\rm cv}(x) &=& \sum_{w,z}\bar q(w) t^a K_\mu(w,z|x)q(z),\\
  A_\mu^{a,\rm cv}(x) &=& \sum_{w,z}\bar q(w) t^a K_\mu(w,z|x)[\hat\gamma_5 q](z),
\end{eqnarray}
where $t^a$ denotes the generator of $SU(N_f)$ normalized as
${\rm Tr}\,t^at^b=\delta^{ab}/2$.
For flavor SU(2), $t^a=\tau^a/2$ with $\tau^2$ the Pauli matrix.
In the following, we consider the flavor non-singlet currents.

In practical implementation for numerical calculations, we approximate
the sign function in (\ref{eq:overlap}) 
by a rational function with Zolotarev's optimized coefficients. 
In our setup, the sign function is approximated to the level of
$10^{-(7-8)}$ with the number of pole $N\simeq$ 10.
(For details, see \cite{Aoki:2008tq} for instance.)
Accordingly, the kernel $K_\mu(w,z|x)$ is constructed as
\begin{eqnarray}
  \label{eq:kernel}
  K_\mu(w,z|x) &=& m_0\left(1-\frac{m}{2m_0}\right)\gamma_5
  \left[ 
   \frac{d_0}{\lambda_{\rm  min}}W_\mu(h_W^2+c_{2n})
   \sum_{l=1}^N \frac{b_l}{h_W^2+c_{2l-1}} 
   \right.
   \nonumber\\
  &+&  
  \left.
    \frac{d_0}{\lambda_{\rm min}}h_W
    \sum_{l=1}^N (c_{2l-1}-c_{2n})\frac{b_l}{h_W^2+c_{2l-1}}
    (W_\mu h_W+h_W W_\mu)\frac{1}{h_W^2+c_{2l-1}} 
  \right],
\end{eqnarray}
with 
\begin{eqnarray}
  W_\mu(z,w|x) &=& -\frac{1}{2}\gamma_5
  \left\{
    (1-\gamma_\mu)U_\mu(z)\delta_{x+\hat\mu,w}\delta_{x,z}
    -
    (1+\gamma_\mu)U^\dag_\mu(z-\hat\mu)\delta_{x,w}\delta_{z-\hat\mu,x}
  \right\},
\end{eqnarray}
and
$h_W(w,z)=H_W(w,z;-m_0)/\lambda_{\rm min}$, where $\lambda_{\rm min}$ is a
lower limit of the eigenvalue of $|H_W|$ to be approximated by the
rational function. 
(In (\ref{eq:kernel}) the site indices of $W_\mu$ and $h_W$ are
omitted, but they are multiplied as matrices.)
The Zolotarev's coefficients $b_l$, $c_l$, $d_0$ are given in 
\cite{Aoki:2008tq}.

In this study, we consider the two-point functions of
flavor non-singlet conserved and local currents
$\langle J_\mu^{a,\rm cv}(x) J_\nu^{b,\rm loc}(0)\rangle$, where
$J_\mu^{a,\rm loc}(x)$ is either
$V_\mu^{a,\rm loc}(x)=Z\bar q(x) t^a \gamma_\mu q(x)$ or
$A_\mu^{a,\rm loc}(x)=Z\bar q(x) t^a \gamma_\mu\gamma_5 q(x)$.
The renormalization constant $Z$ needed for the local currents to match
their continuum counterpart is non-perturbatively determined as
$Z=1.39360(48)$ \cite{Noaki:2009xi}.  
The lattice calculation of 
$\langle J_\mu^{a,\rm cv}(x) J_\nu^{b,\rm loc}(y)\rangle$
is standard except for the complicated form of $J_\mu^{a,\rm cv}(x)$.
Namely, we calculate the quark propagator originating from a
fixed space-time point $y=0$ and construct the two-point function with
the conserved current $J_\mu^{a,\rm cv}(x)$ located at arbitrary space-time
point $x$. 
We then apply the Fourier transform in all four dimensions to obtain
the two-point function in the momentum space.

Because of the current conservation of $J_\mu^{a,\rm cv}$, we may derive
the WT identities for the two-point functions
\begin{eqnarray}
&&\sum_\mu \hat Q_\mu\langle V_\mu^{a,\rm cv} V_\nu^{b,\rm loc}\rangle(Q) = 0, \label{eq:VWT}\\
&&\sum_\mu \hat Q_\mu\langle A_\mu^{a,\rm cv} A_\nu^{b,\rm loc}\rangle(Q) 
  - 2m_q\langle P^a A_\nu^{b,\rm loc}\rangle(Q) 
  = 0,
\label{eq:AWT}
\end{eqnarray}
where $a\hat Q_\mu = \sin(aQ_\mu)$ are a momentum definition
corresponding to the backward derivative operator $\partial_\mu^{x\,*}$.
We use a convention that the two-point function after the
Fourier transformation, such as 
$\langle J^a_\mu J^b_\nu\rangle(Q)$, 
is a function of $aQ_\mu=2\pi n_\mu/L_\mu$ 
with $L_{\mu=1\sim 4}$ the extent of the lattice in the $\mu$-th direction.
The second term in (\ref{eq:AWT}) represents the correlation function
of the pseudo-scalar density operator  
$P^a(x)=\bar q(x)t^a\gamma_5(1-D_{\rm ov}/m_0)q(x)$ and the local axial-vector
current $A_\nu^{b,\rm loc}(y)$. 
A possible term arising from the axial transformation of 
$J_\nu^{b,\rm loc}(y)$ ($J=V$ or $A$) 
vanishes when we take the vacuum expectation value,
since the vacuum has axis-interchange symmetry while the index $\nu$
remains in $J_\nu^{b,\rm loc}(y)$.

The vector and axial-vector VPFs are now given by
\begin{eqnarray}
  \langle J^{a,\rm cv}_\mu J^{b,\rm loc}_\nu\rangle(Q) &=& 
  \delta^{ab} \left[
    (\delta_{\mu\nu}\hat Q^2-\hat Q_\mu \hat Q_\nu)\Pi_J^{(1)}(Q)
    - \hat Q_\mu \hat Q_\nu \Pi_J^{(0)}(Q) + \Delta^J_{\mu\nu}(Q)
  \right].
  \label{eq:JcvJloc}
\label{eq:JJlat}
\end{eqnarray}
Here, $\Pi_V^{(0)}(Q)$ vanishes because of the conservation of
$V_\mu^{a,\rm cv}$, while 
$\Pi_A^{(0)}(Q)$ represents a remnant due to PCAC:
\begin{equation}
  \label{eq:Pi0}
  \Pi_A^{(0)}(Q) = 
  -2m_q\langle P^a A_\nu^{a,\rm loc}\rangle(Q)/(\hat Q^2 \hat Q_\nu).
\end{equation}
(Repeated indices $a$'s are not summed.)
The transverse part $\Pi_J^{(1)}(Q)$ can be extracted as
\begin{equation}
  \label{eq:Pi1}
  \Pi_J^{(1)}(Q) = \langle J^{a,\rm cv}_\mu J^{a,\rm loc}_\mu\rangle(Q)
  / (\hat Q^2-\hat Q_\mu\hat Q_\mu),
\end{equation}
(repeated indices $\mu$'s are not summed)
if one ignores the additional term $\Delta_{\mu\nu}^J(Q)$, which
reflects the violation of the current conservation of the local
current $J_\nu^{a,\rm loc}$.
Since the current conservation is recovered in the continuum limit,
this term can be expanded in terms of small $aQ_\mu$ as
\begin{eqnarray}
  \Delta_{\mu\nu}^J(Q) &=& \sum_{m,n=1}\Big(\delta_{\mu\nu}\sum_\rho 
  \hat Q_\rho^{2m} - \hat Q_\mu^{2(m-1)}\hat Q_\mu \hat Q_\nu\Big)Q_\nu^{2n}
   F_{mn}(\hat Q).
  \label{eq:delta}
\end{eqnarray}
where $F_{mn}$ denotes the scalar function depends on the index $m,n$ and momentum $Q$.
It satisfies the condition
$\sum_\mu \hat Q_\mu\Delta_{\mu\nu}^J(Q)=0$ coming from 
the WT identity for $J_\mu^{a,\rm cv}$. 
In this work, we confirmed that this term is numerically negligible in the 
range $(aQ)^2<1$, and ignore its contribution as we discuss later.

\section{Fit with the perturbative formula}
\label{sec:fit}

Defining $\Pi_J(Q)=\Pi_J^{(0)}(Q)+\Pi_J^{(1)}(Q)$, 
the Operator Product Expansion (OPE) of VPF,
$\Pi_{V+A}(Q)=\Pi_V(Q)+\Pi_A(Q)$,
is given by
\begin{eqnarray}
 \Pi_{V+A}|_{\rm OPE}(Q^2,\alpha_s) &=& c + C_0(Q^2,\mu^2,\alpha_s)\nonumber\\
 &+& C_m^{V+A}(Q^2,\mu^2,\alpha_s)\frac{\bar{m}^2(Q)}{Q^2}\nonumber\\
 &+& \sum_{q=u,d,s} C^{V+A}_{\bar qq}(Q^2,\alpha_s)\frac{\langle m_q\bar q q\rangle}{Q^4}\nonumber\\
 &+& C_{GG}(Q^2,\alpha_s)\frac{\langle(\alpha_s/\pi) GG\rangle}{Q^4}
    + \mathcal O(Q^{-6})
  \label{eq:pi_J_OPE}
\end{eqnarray}
for large $Q^2$.
The perturbative expansion of the coefficients $C_X^{(V+A)}$ 
($X=0$, $\bar qq$ and $GG$) is known up to two- to four-loop order
in the continuum renormalization scheme, {\it i.e.} the
$\overline{\mathrm{MS}}$ scheme, depending on the terms. 

The first term $c$ in (\ref{eq:pi_J_OPE}) is a scheme-dependent
constant, divergent in the limit of infinite ultraviolet cutoff.
For the Adler function $D(Q^2)=-Q^2 d\Pi(Q^2)/dQ^2$, which
is a physical observable, the first term disappears and the contributions
from other terms remain finite.
The coefficients in the second and third terms are perturbatively
calculated to four-loop order in the $\overline{\mathrm{MS}}$ scheme
\cite{Surguladze:1990tg,Gorishnii:1990vf,Chetyrkin:1996};
the expression explicitly contains $\alpha_s^{(3)}(Q)$ defined in the
$\overline{\mathrm{MS}}$ scheme. 
(The superscript (3) stands for the number of flavors.)
The third term contains the running mass $\bar{m}(Q)$ whose
anomalous dimension is known to three-loop order
\cite{Chetyrkin:1985kn,Chetyrkin:1993}. 
The fourth and fifth terms represent higher order effects in OPE
containing dimension-four operators.
Their Wilson coefficients are calculated at three-loop order
\cite{Chetyrkin:1985kn}.

In addition to the terms represented by the continuum OPE
(\ref{eq:pi_J_OPE}), 
there are discretization effects of $\mathcal O(a^2Q^2)$ at finite
lattice spacings.
These can be eliminated by an extrapolation to the continuum limit, in
principle.  
In our calculation obtained at a single lattice spacing, however, 
the error has to be carefully investigated. 
We use a lattice perturbation theory to estimate the discretization
effects at large $a^2 Q^2$ regime as described below.
We also note that the exact symmetries of the overlap fermion partly
eliminate unphysical terms of $\mathcal O(a^2Q^2)$ that 
violate the WT identities \cite{Shintani:2008ga}. 
We therefore use (\ref{eq:pi_J_OPE}) without including correction
terms describing the discretization effects when we fit the lattice data
of VPF extracted through (\ref{eq:Pi0}) and (\ref{eq:Pi1}).
In our previous study in two-flavor QCD \cite{Shintani:2008ga}, we had
to use more complicated method to extract the physical VPFs, because
of non-conserved (axial-)vector currents. 

We now discuss a fit of the lattice VPF data to the OPE formula.
In this analysis the renormalization scale is set to $\mu=2$~GeV when
necessary, though the final result should not depend on $\mu$ up to
higher order perturbative corrections. 
The gluon condensate $\langle(\alpha_s/\pi)GG\rangle$ is defined only
through the perturbative expression like (\ref{eq:pi_J_OPE}) because
of the renormalon ambiguity \cite{Martinelli:1996pk}, 
hence we treat
$\langle(\alpha_s/\pi)GG\rangle$ as a free parameter to describe 
associated $1/Q^4$ corrections. 
On the other hand, the quark condensate $\langle \bar{q}q\rangle$ is
well-defined in the massless limit, as there is no mixing with
lower dimensional operators because of the exact chiral symmetry of
overlap fermion.  
Thus, the quark mass dependence of 
$\Pi_{V+A}|_{\rm OPE}(Q^2,\alpha_s)$, which comes from the third and
fourth terms in (\ref{eq:pi_J_OPE}), is given only as a function
of $\alpha_s$, once the quark condensate is determined elsewhere.

The running quark mass $\bar{m}(Q)$ is set to the value
corresponding to the quark mass used in the lattice calculation.
First, we obtain the value at 2~GeV using the non-perturbatively
calculated $Z$-factor as
$\bar{m}(2\mathrm{~GeV})= Z_m({\rm 2\,GeV})m_q$
with $Z_m({\rm 2\,GeV})$ = 0.806(12)(24)($^{+\ 0}_{-11}$) 
\cite{Noaki:2008iy}.
Then, it is evolved to $Q^2$ using a three-loop running formula
\cite{Chetyrkin:1985kn,Chetyrkin:1993}.

For the quark condensate of up and down quarks, we use the value
obtained in the recent analysis of the spectral density
\cite{Fukaya:2009fh}, 
{\it i.e.} 
$\langle\bar qq\rangle$ = $-$[0.242(04)($^{+19}_{-18}$) GeV]$^3$,
which is defined in the $\overline{\mathrm{MS}}$ scheme at 2 GeV.
The strange quark condensate $\langle\bar ss\rangle$ appears only as a
contribution from sea quark and the associated coefficient
$C_{\bar{s}s}^{V+A}(Q^2,\alpha_s)$ starts from $O(\alpha_s)$.
For the value of $\langle\bar ss\rangle$, we use the same value as the
one of up and down quarks. 

In the fit of VPF using (\ref{eq:pi_J_OPE}), there are three unknown
parameters, $\alpha_s$, $c$ and $\langle(\alpha_s/\pi)GG\rangle$.
The QCD scale $\Lambda^{(3)}_{\overline{MS}}$ controls the running
coupling constant $\alpha_s^{(3)}(Q)$, which is evaluated using the
four-loop formula \cite{van Ritbergen:1997va,Czakon:2004bu}. 

\begin{figure}[p]
  \begin{center}
    \includegraphics[width=140mm]{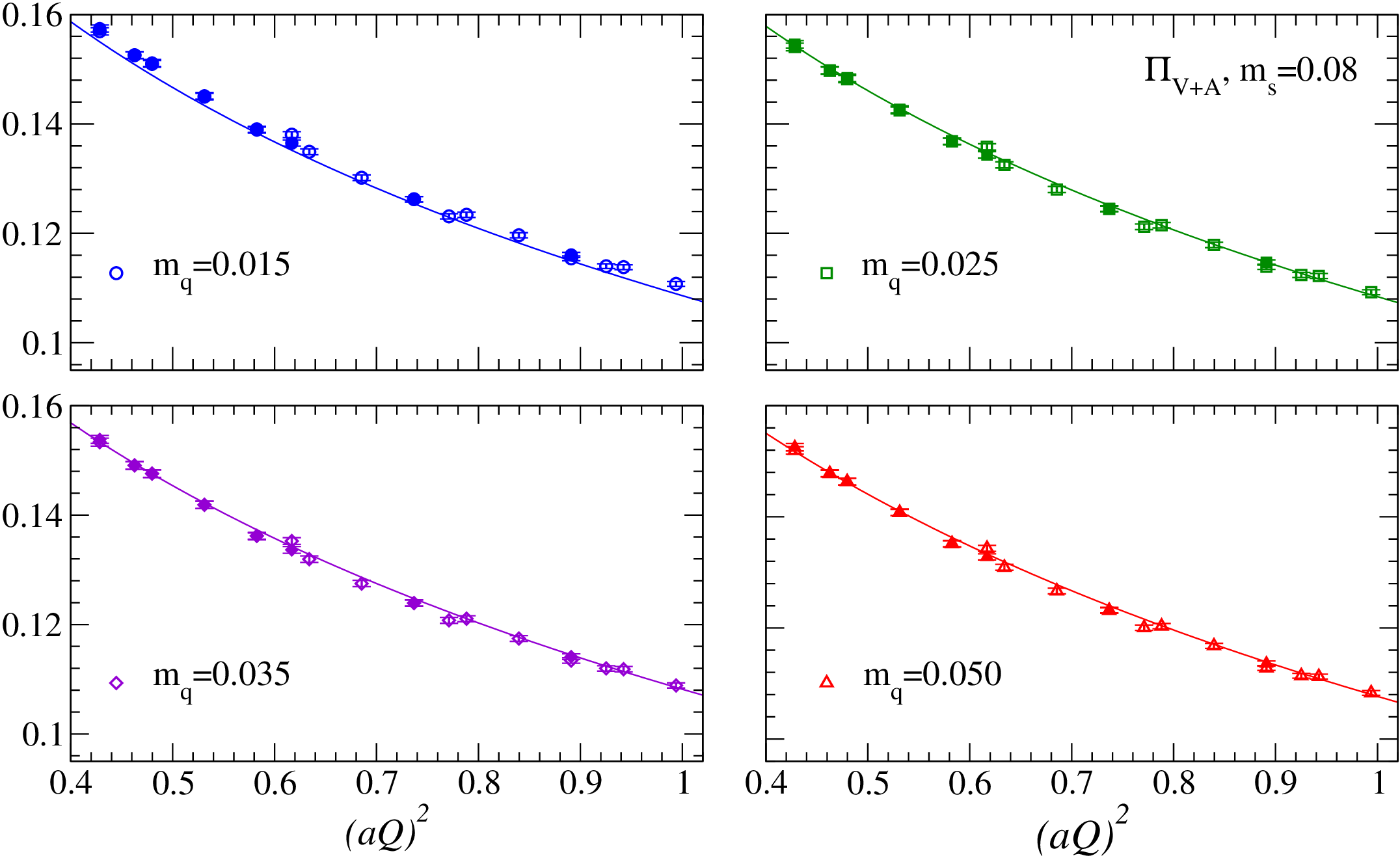}
    \vskip 5mm
    \includegraphics[width=140mm]{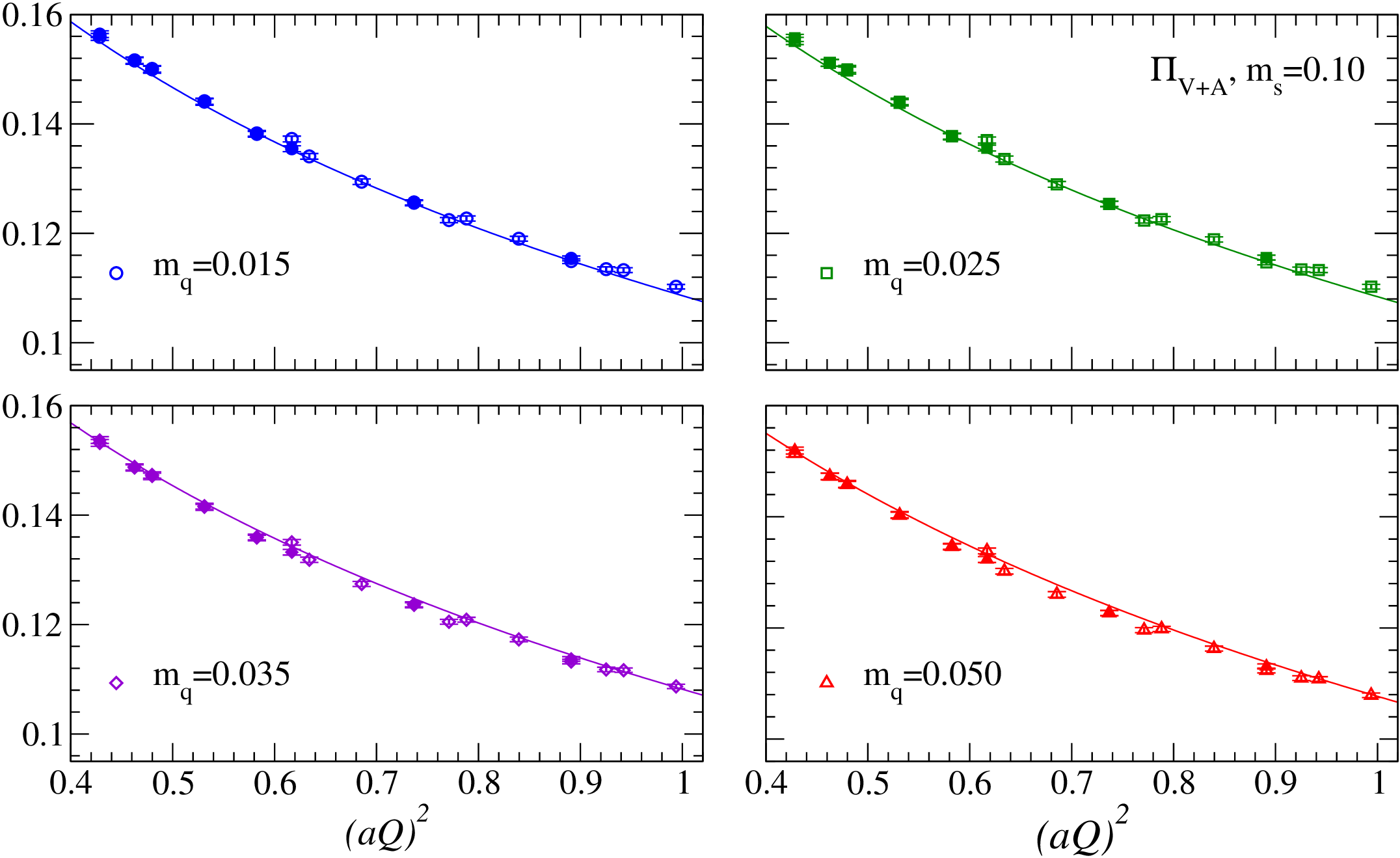}
    \caption{
    $(aQ)^2$ dependence of VPF, $\Pi_{V+A}(Q)$, at all valence quark
    masses: 
    $m_q=0.015$ (circle), $0.025$ (square), $0.035$ (diamond), and
    $0.050$ (triangle).
    Top half is a result at $m_s=0.08$ while the bottom is at $m_s=0.10$.
    Solid curves show a fit function at each quark masses.
    Filled symbols are the points for which each momentum component is
    equal to or smaller than $2\pi/16$ in the lattice unit.
    }
\label{fig:PiV+A}
\end{center}
\end{figure}

\begin{figure}
  \begin{center}
    \includegraphics[width=100mm]{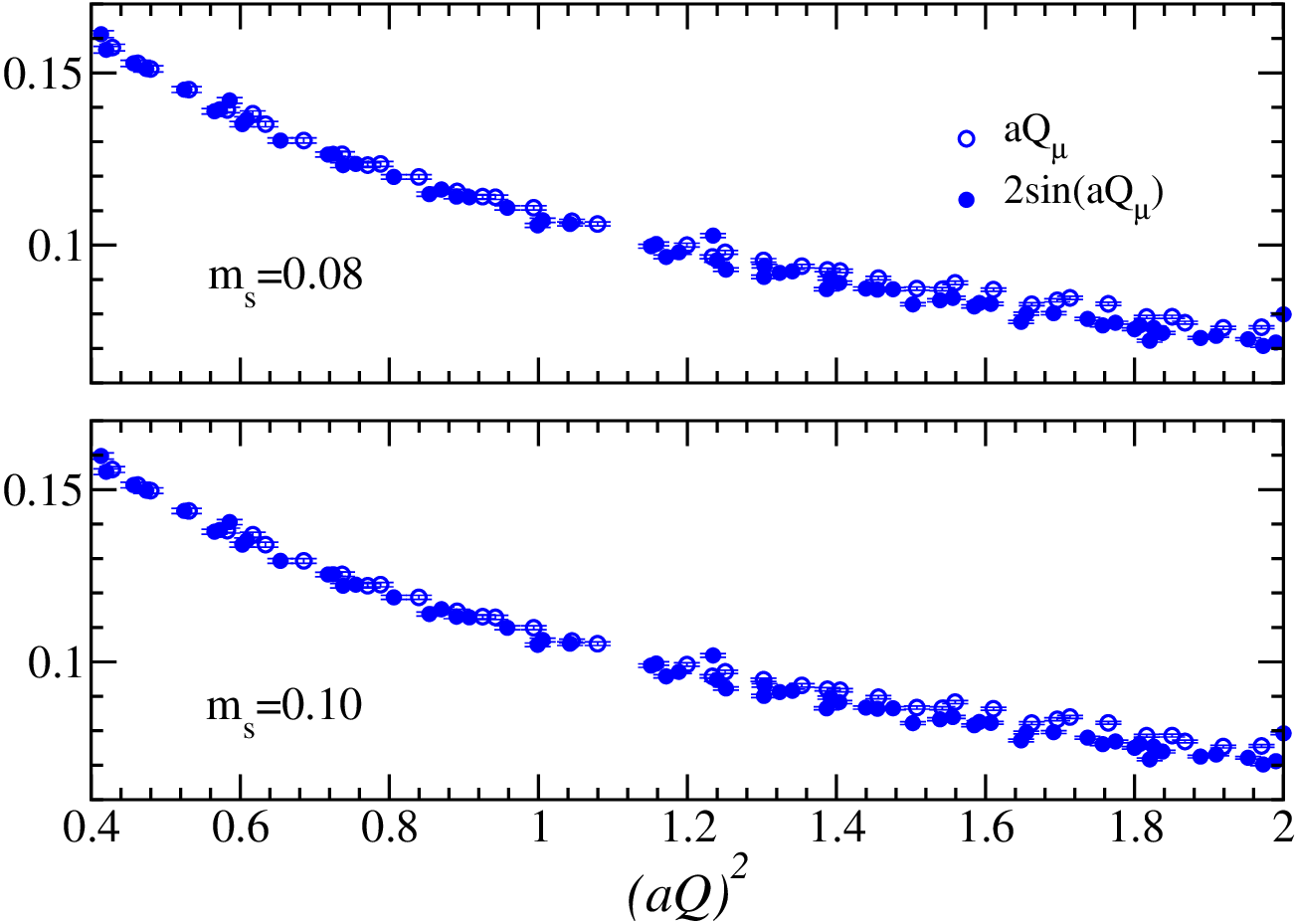}
    \caption{
      Comparison of $\Pi_{V+A}(Q)$ with different momentum
      definitions. 
      Lattice data at $m_q=0.015$.
    }
\label{fig:diff_pdef}
\end{center}
\end{figure}

Figure~\ref{fig:PiV+A} shows a $(aQ)^2$ dependence of $\Pi_{V+A}(Q)$
in a window $0.4\le (aQ)^2\le 1.0$.  
Fit curves shown in this plot are those of (\ref{eq:pi_J_OPE}) with
the value of parameters extracted from the fit in the range
$0.463\le (aQ)^2\le 0.994$.
The upper limit of the range is chosen to avoid significant lattice
artifact, which is estimated by a difference of the lattice momentum
$aQ_\mu$ from the other definition $a\hat Q_\mu$.  
In fact, the result is unchanged within $1\sigma$ level when we use
these different definitions of the momentum as
Figure~\ref{fig:diff_pdef} shows, as far as $(aQ)^2$ is lower than 1.0.
Beyond this value we observe significant deviations between the
different definitions.
We also impose a constraint $aQ_\mu < \pi/4$ for each momentum
component to avoid large lattice artifacts.

\begin{figure}
  \begin{center}
    \includegraphics[width=110mm]{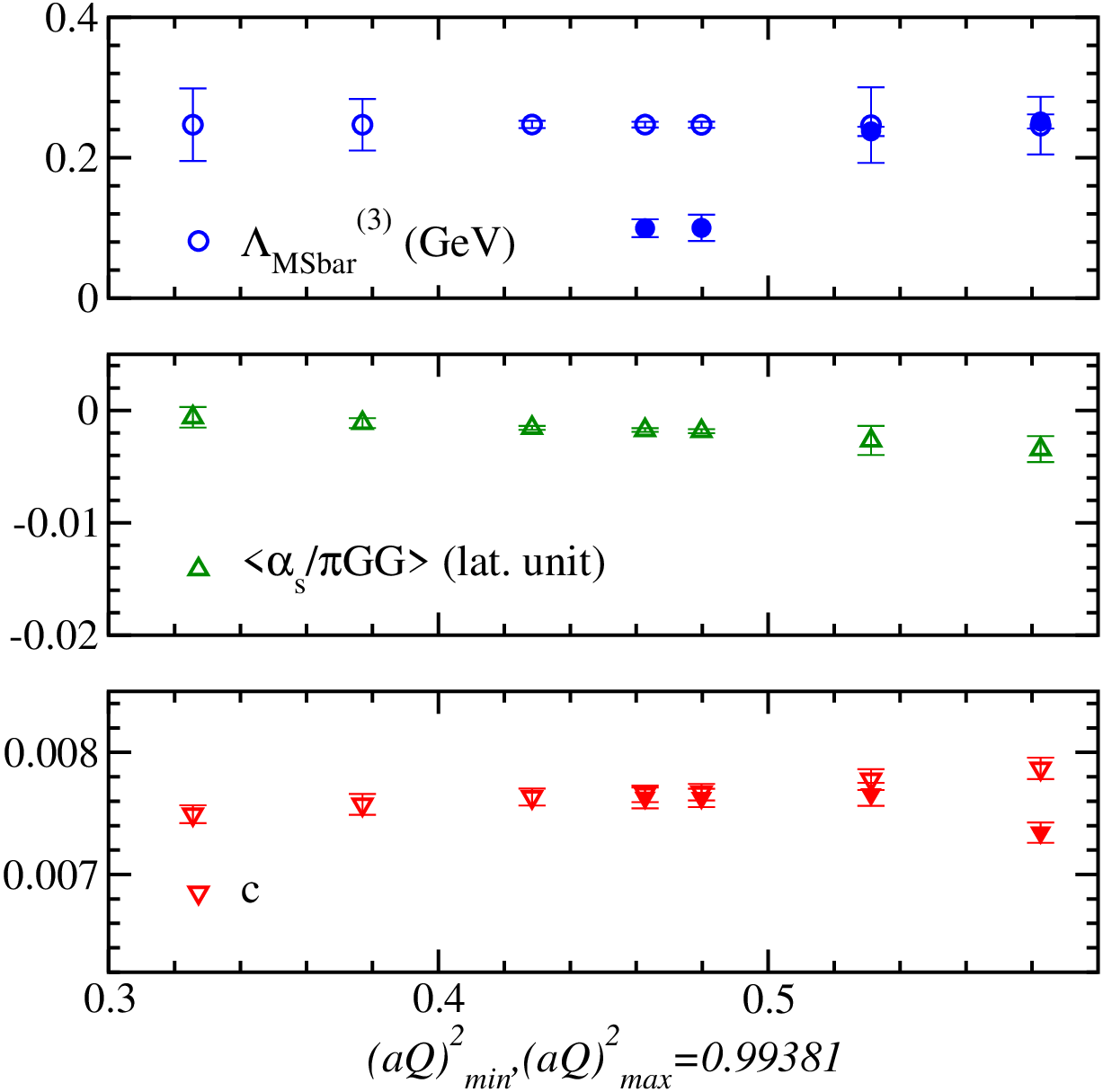}
    \caption{
    Dependence of the fit parameters on the lower limit of the fit range.
    The maximum value is fixed at $(aQ)^2\simeq 0.994$.
    Open and filled symbols show the results with and without the
    $1/Q^4$ terms in (\ref{eq:pi_J_OPE}).  
    (Thus, there is no filled symbol in the middle plot.)
    }
\label{fig:fitdep}
\end{center}
\end{figure}

In order to determine the lower limit, we investigate the stability of
the fit results. 
Figure~\ref{fig:fitdep} shows the dependence of fit parameters on the
value of the lower limit $(aQ)^2_{\rm min}$.
We observe that around $(aQ)^2_{\rm min}=$ 0.4--0.5 all the parameters
are stable. 

It is interesting to see where the $1/Q^4$ term becomes significant.
In Figure~\ref{fig:fitdep}, the fit results without the $1/Q^4$ terms
are also shown by filled symbols.
It turned out that the value of
$\Lambda_{\overline{\mathrm{MS}}}^{(3)}$ 
is consistent with the $1/Q^4$ fits when $(aQ)^2_{\rm min}$ is greater
than 0.5, which means that the $1/Q^4$ terms become relevant below
this value.
In fact, if we extend this fit including the data points slightly below
$(aQ)^2=0.5$, the value of $\Lambda_{\overline{\mathrm{MS}}}^{(3)}$
becomes significantly lower;
the $\chi^2/{\rm dof}$ of the fit becomes too large ($\sim$ 3.0),
which suggests that the fit is no longer valid.
Strictly speaking, $\chi^2/{\rm dof}$ does not have a statistical
meaning as the correlation among the data at difference $Q^2$ is
ignored in the fit used here.  
We discuss on the statistical correlations among the data points 
in the next section.

\begin{figure}
  \begin{center}
    \includegraphics[width=110mm]{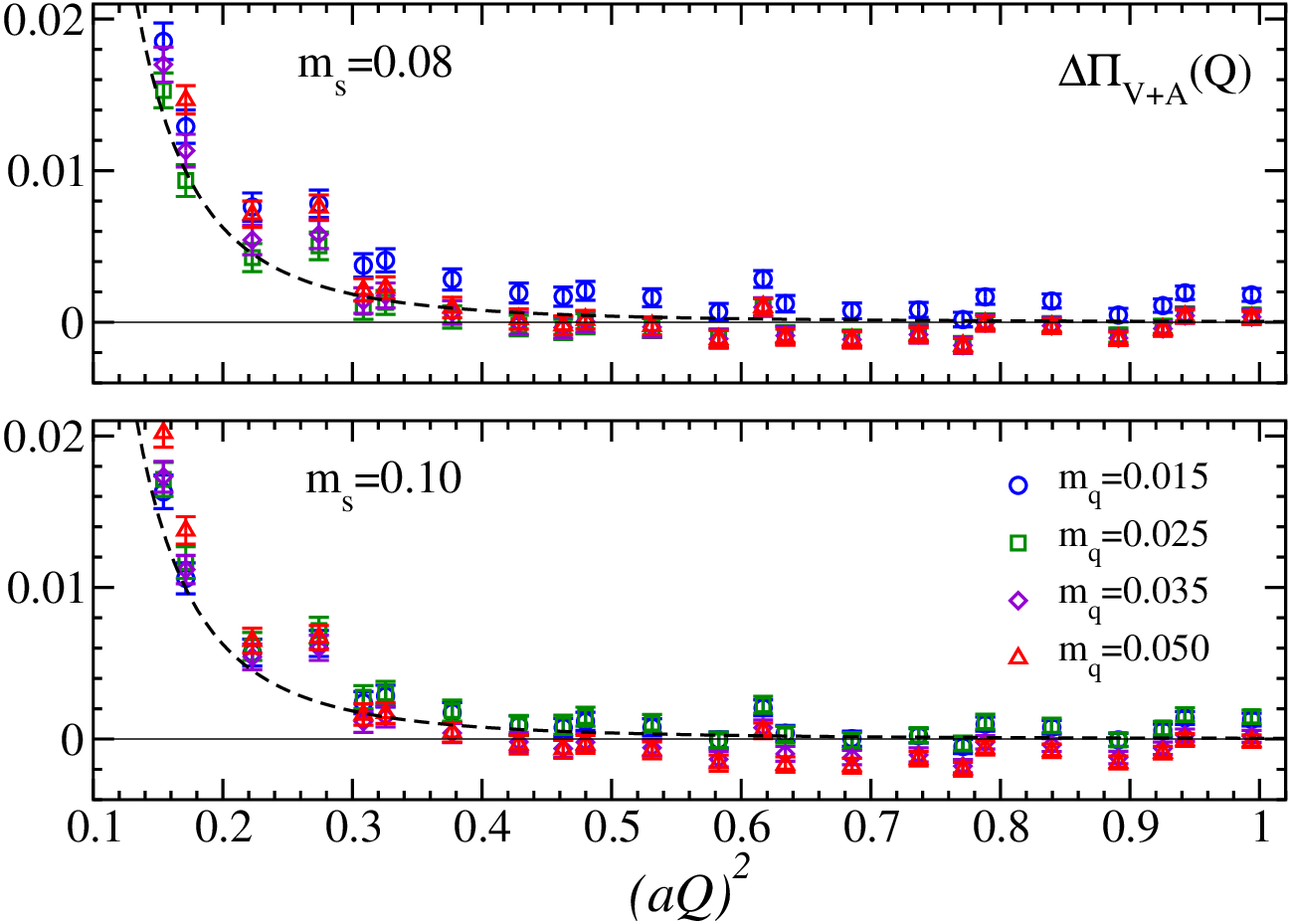}
    \caption{
    Difference between the lattice data and the fit function
    (\ref{eq:pi_J_OPE}). 
    Dashed line shows a guiding line representing the $1/Q^6$
    behavior. 
    }
\label{fig:diffPiV+A}
\end{center}
\end{figure}

The limitation of the OPE formula including up to $1/Q^4$ terms can be
investigated by looking at its departure from the lattice data at
lower values of $Q^2$.
In Figure~\ref{fig:diffPiV+A} we plot the difference of the lattice
data from the fit curve including the $1/Q^4$ terms.
The $Q^2$ region in this plot is extended towards the value below
$(aQ)^2_{\rm min}$.
From this plot, we observe that the next order $1/Q^6$ contribution
becomes significant below $(aQ)^2\simeq$ 0.4.
We therefore set $(aQ)^2_{\rm min}= 0.463$ in our analysis including
the $1/Q^4$ terms.

After doing a simultaneous fit of the VPF data at all sea quark
masses, the QCD parameter is obtained as 
$\Lambda_{\overline{MS}}^{(3)}$ = 0.247(5)~GeV.
By matching onto four- and then to five-flavor QCD at charm and bottom
quark masses respectively, 
the strong coupling constant is obtained
as $\alpha_s^{(5)}(M_Z)=0.1181(3)$
at the $Z$ boson mass scale.
Here, the error is statistical only.
Various sources of the systematic error are discussed in the next
section.

\section{Systematic errors and final result}
\label{sec:error}

\subsection{Uncorrelated fit}
First of all, our fit procedure of VPF may induce systematic error
due to the use of uncorrelated fit.
Namely, in the fit described above, we did not take the correlation
among different $Q^2$ points into account and estimated the
statistical error for the fit parameters using the jackknife method.

In order to estimate the associated error, we calculated the
statistical correlation of different $Q^2$ points and found it very
strong (50--100\%). 
If we construct $\chi^2$ taking account of the correlation, the value
of $\chi^2$/dof is of order 100.
This unacceptably large value occurs because the fit function
(\ref{eq:pi_J_OPE})  does not contain the discretization effects that
violate Lorentz symmetry.
Indeed, if we restrict the data points for those that each momentum
component is equal to or smaller than $2\pi/16$,
the $\chi^2$/dof is reduced to 1.7, without changing the central
values of the fit parameters. 
The restricted data points are shown in Figure~\ref{fig:PiV+A} by
filled symbols. 

In the main analysis we use all the data points that satisfy the
condition $aQ_\mu < 4\pi/16$ in order to use as much information from
the lattice data as possible with the uncorrelated fit.
In particular, we can take a wider range of $Q^2$ with this choice,
that improves the stability of the fit.
In other words, with the limited data points ($Q_\mu\le 2\pi/16$) the
$\chi^2$ fit is sometimes trapped in a local minimum depending on the
initial values for the fit parameters.

We therefore decided to use the uncorrelated fit for the enlarged data
points ($Q_\mu < 4\pi/16$) to obtain the fit parameters, 
and then to check the value of $\chi^2/\mathrm{dof}$ for the limited
data points ($Q_\mu\le 2\pi/16$) taking account of the correlation.
Since the value of $\chi^2/\mathrm{dof}$ is 1.7, we do not expect
the bias due to this procedure larger than one standard deviation,
assuming that the full correlated fit should give 
$\chi^2/\mathrm{dof}\sim 1$.
Thus, we conservatively assign a systematic error $\pm 0.003$ for
$\alpha_s^{(5)}(M_Z)$, which is equal to the size of the statistical
error of one standard deviation.

This procedure can be avoided if the lattice data are obtained at
finer lattice spacings so that one can cover the same physical range of
$Q^2$ with smaller lattice momenta.

\subsection{Discretization effect}
As described above, the discretization effect is significant in our
lattice data especially when we try to cover large enough $Q^2$ range.
We estimate the associated error using lattice perturbation theory.

Since the discretization effect is most important in the large
momentum region, the perturbation theory can be used to estimate its
size. 
We calculate the one-loop diagram of VPF, $\Pi_{V+A}^{\rm PT}(Q)$,
with local and conserved currents in lattice perturbation theory, and
compare them with the continuum perturbation theory.
This provides an estimate of the discretization effect at the zeroth
order of $\alpha_s$.
Because the discretization error itself is a small effect, its
calculation at the leading order gives a reasonably precise estimate.

\begin{figure}
  \begin{center}
    \includegraphics[width=100mm]{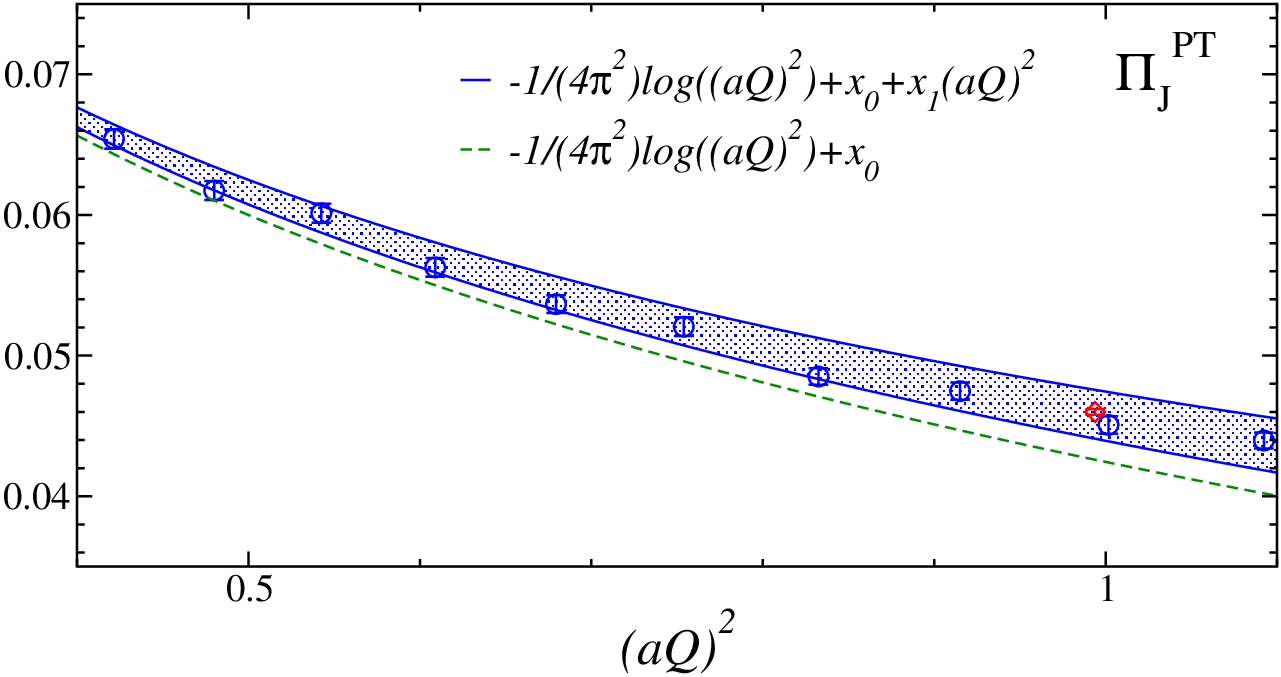}
    \caption{
      $(aQ)^2$ dependence of one-loop VPF $\Pi_{J=V,A}(Q^2)$ in lattice
      perturbation theory. 
      Dashed line shows the leading logarithm term plus a constant,
      which corresponds to the continuum perturbation theory.
      Solid lines show the function including lattice artifact of
      $O((aQ^2))$.  
      The shaded band represents an uncertainty due to the higher
      order effects.
      The red diamond denotes the value at the upper limit of our fit
      of VPF.
    }
\label{fig:latPT}
\end{center}
\end{figure}

Figure~\ref{fig:latPT} shows the result as a function of $(aQ)^2$.
Since the lattice regularization violates rotational invariance, the
result is not a completely smooth function of $(aQ)^2$, as shown by
squares in the plot, which correspond to representative points of
$(aQ)^2$ in our lattice calculation.

The perturbative result may be parametrized as 
$\Pi_{V+A}(Q^2)^{\rm latt.pert}
=c-1/(2\pi^2)\ln(aQ)^2+0.0062(40)(aQ)^2$ for small $a$.
The logarithmic term is the same as in the continuum perturbation
theory and $c$ is the scheme dependent constant as already noticed.
The term $+0.0062(40)(aQ)^2$ comes from the discretization effect.
The error includes a fluctuation of numerical integral as well as the
non-smooth behavior due to higher order discretization effects. 
The non-smooth behavior appears because of different assignments of
momentum components $aQ_\mu$.
In the plots we took several values of $(aQ)^2$ (and so $aQ_\mu$) that
also appear in the lattice calculation.
We observe that the pattern of non-smooth behavior in the one-loop
calculation actually well reproduces that occurring in the numerical
simulation.
It suggests that our estimate of the discretization effect is
reasonably realistic.
As the plot shows, this error band is taken so that the result of the
lattice perturbation theory calculation is covered.

By subtracting this estimate of the $\mathcal{O}((aQ)^2)$ effect from the
lattice data,
the final result for $\alpha^{(5)}(M_Z)$ changes by $+0.0002(1)$.
We therefore take $+0.0003$ as our estimate of the systematic error
from this source, without changing the central value to be conservative.
The estimated error in the negative direction is thus taken to be zero.
Although the perturbative calculation is done only at the one-loop
level, we expect that the higher order effects are suppressed by an
additional factor of $\alpha_s$ and thus well below $\pm 0.0001$.

\subsection{Non-conserved current}
The discretization effect may also come from the non-conserved local
current $J_\nu^{\rm loc}$ in (\ref{eq:delta}), which is represented by
the term $\Delta_{\mu\nu}^J(Q)$ in (\ref{eq:JcvJloc}).

We estimate its leading contribution
$\Delta^J_{\mu\nu}(Q)=(\delta_{\mu\nu}\hat Q^2-\hat Q_\mu\hat Q_\nu)
Q_\nu^2 F^J_{11}(Q)$
in the expansion (\ref{eq:delta}) in terms of small $aQ_\mu$,
by solving linear equations (\ref{eq:JcvJloc}) for different sets of
$\mu$ and $\nu$.  
We find that the maximum magnitude of 
$(\delta_{\mu\nu}\hat Q^2-\hat Q_\mu\hat Q_\nu) Q_\nu^2 F^{V+A}_{11}(Q)$ 
is less than 1\% of $\Pi_{V+A}(Q)$ in the fit range 
$0.463\le(aQ)^2\le 0.994$. 

\begin{figure}
\begin{center}
  \includegraphics[width=100mm]{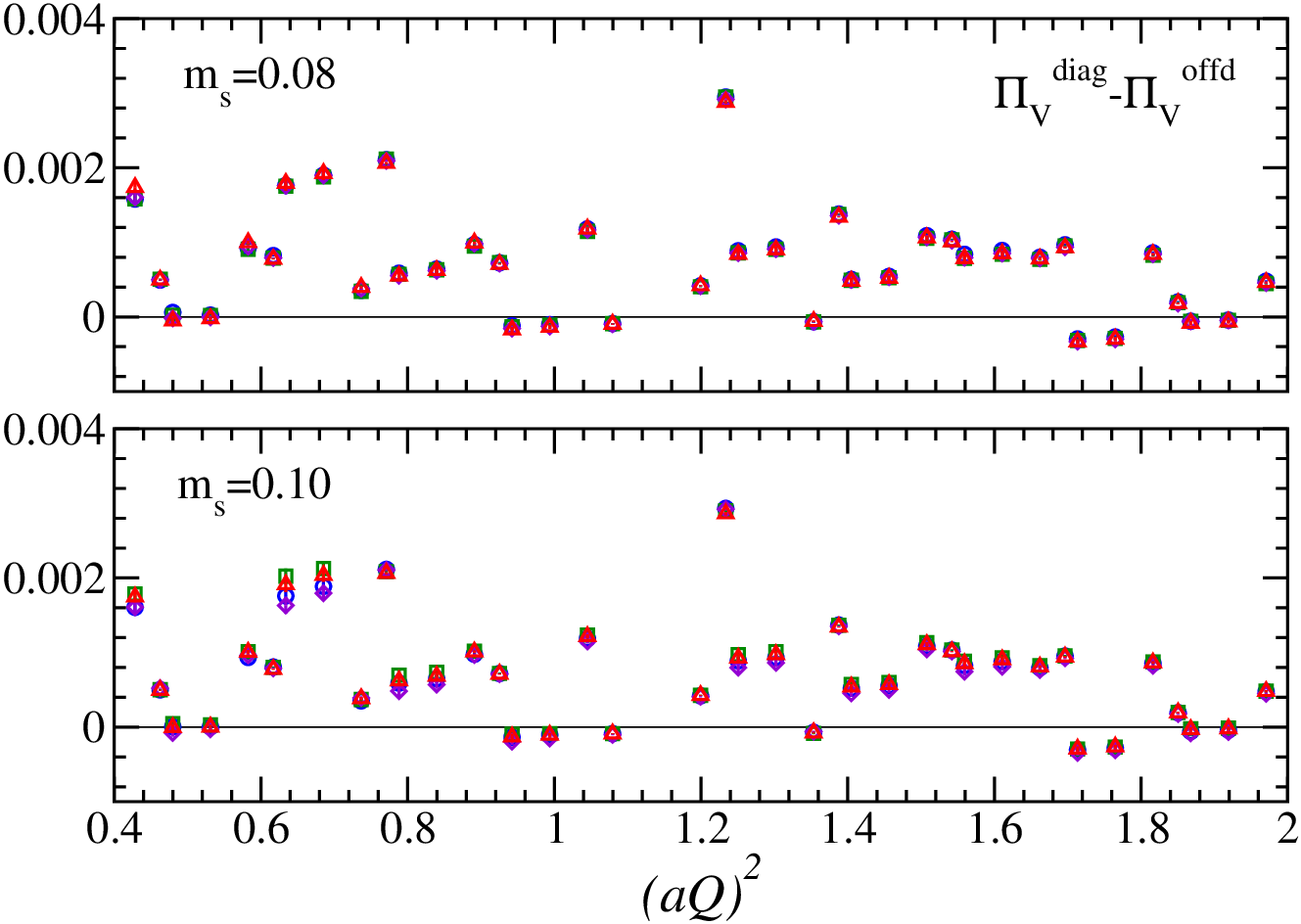}
  \caption{
    Difference between $\Pi_V^{\rm diag}(Q)$ and $\Pi_V^{\rm offd}(Q)$ at
    all valence quark masses 
    $m_q=0.015$ (circles), $0.025$ (squares), $0.035$ (diamonds), and
    $0.050$ (triangles).  
    Top panel is the result at $m_s=0.08$ and the bottom is in $m_s=0.10$.
    }
  \label{fig:diff_diod}
\end{center}
\end{figure}

The contribution of $\Delta_{\mu\nu}^J(Q)$ may also be estimated by
looking at a difference between VPFs obtained with $\mu=\nu$ (diag) and 
with $\mu\not=\nu$ (offd) components, {\it i.e.}
\begin{eqnarray}
\Pi_J^{\rm diag}(Q) &=& \langle J^{\rm cv}_\mu J^{\rm loc}_\mu\rangle(Q)
                 / (\hat Q^2-\hat Q_\mu\hat Q_\mu),\\
\Pi_J^{\rm offd}(Q) &=& \langle J^{\rm cv}_\mu J^{\rm loc}_\nu\rangle(Q)
                 / (-\hat Q_\mu\hat Q_\nu),
\end{eqnarray}
respectively.
Figure \ref{fig:diff_diod} shows 
$\Pi_V^{\rm diag}(Q)-\Pi_V^{\rm offd}(Q)$ 
as a function of $(aQ)^2$.
The maximum magnitude of the difference in the range
$0.463\le (aQ)^2\le 0.994$ is 0.003, which is the same order as the
estimate from $F^{V+A}_{11}(Q)$.  

Adding or subtracting this amount of systematic effect from the
lattice data, we repeat the whole analysis to estimate the systematic
error due to the Lorentz (or WT) violating terms, which gives
$\pm 0.0002$ in $\alpha_s^{(5)}(M_Z)$.

\subsection{Quark condensate and $Z_m$}
The uncertainty due to the quark condensate is estimated as 
$\pm 0.0001$ for $\alpha_s^{(5)}(M_Z)$ 
by varying the input value from $-[0.220\,{\rm GeV}]^3$ to
$-[0.265\,{\rm GeV}]^3$, which correspond to the lower and upper
limits of the estimate of $\langle\bar{q}q\rangle$ in \cite{Fukaya:2009fh}.

The uncertainty due to $Z_m$ is also estimated as $\pm0.0001$
for $\alpha_s^{(5)}(M_Z)$ by varying $Z_m$ within its estimated error
(from 0.777 to 0.832). 


\subsection{Perturbative expansion}
The truncation effect of the perturbative expansion can be estimated
by comparing the results with different orders of the perturbative
expansion. 
Fortunately, the four-loop calculation is known for
$C_0(Q^2,\mu^2,\alpha_s)$ \cite{Chetyrkin:2006,Boughezal:2006px}, 
and we can explicitly estimate the effect of 
${\mathcal O}(\alpha_s^3)$.

\begin{figure}
  \begin{center}
    \includegraphics[width=100mm]{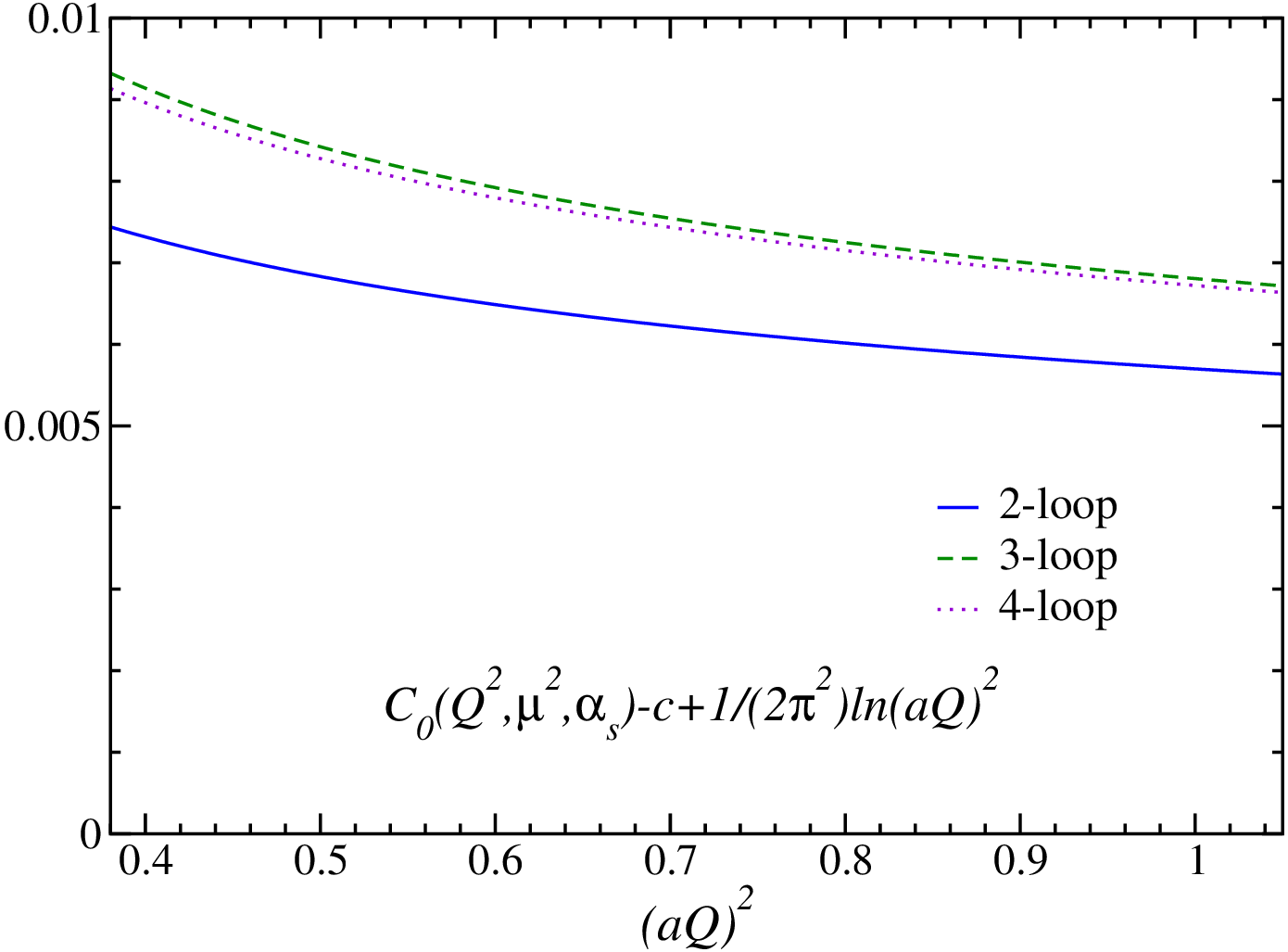}
    \caption{
      $C_0(Q^2,\mu^2,\alpha_s)-c+1/(2\pi^2)\ln(aQ)^2$
      as a function of $(aQ)^2$.
      The perturbative results at two-loop (solid), three-loop
      (dashed), and four-loop (dotted) calculations are shown.
      The logarithm at the leading order is subtracted 
      in order to enhance their small differences.
    }
\label{fig:pi_C0}
\end{center}
\end{figure}

A comparison of two-, three- and four-loop calculations of
$C_0(Q^2,\mu^2,\alpha_s)$ is shown in Figure~\ref{fig:pi_C0}.
They correspond to $\mathcal{O}(\alpha_s)$, 
$\mathcal{O}(\alpha_s^2)$, and $\mathcal{O}(\alpha_s^3)$ calculations,
respectively. 
We observe that the difference between three-loop and four-loop is
of order of 0.0001 for $\Pi_J(Q)$, which is much smaller than other
systematic effect.

Strictly speaking, the smallness of the four-loop contribution does
not guarantee that the unknown higher orders are even smaller.
We therefore attempted to fit the data with a formula including
unknown $\mathcal{O}(\alpha_s^4)$ term $c_0^{(4)}\alpha_s^4(Q)$.
The fit gives $c_0^{(4)}\sim \mathcal{O}(10)$ with a shift of
resulting $\alpha_s^{(5)}(M_Z)$ by $+0.0003$.
We therefore put a conservative systematic error from the truncation
of the perturbation series as $\pm 0.0003$.
This can only be reduced by including the data at higher $Q^2$ values,
which needs finer lattice spacings.

\subsection{$1/Q^2$ expansion}
As previously discussed, the size of neglected $1/Q^6$ terms in the
OPE formula is at most 0.001 for $\Pi_J(Q)$ at the lower end of the
fitting range $(aQ)^2_{\rm min}=0.463$
(see Figure~\ref{fig:diffPiV+A}). 
This is less than 1/5 of the estimated discretization effect discussed
above. 
We therefore expect that the impact on $\alpha_s^{(5)}(M_Z)$ is
smaller than 0.0001.

\subsection{Charm and bottom quark mass}
The uncertainty of charm and bottom quark masses, $m_c$ and $m_b$ used in
a perturbative matching of $\alpha_s^{(3)}$ onto $\alpha_s^{(5)}$ is
$+0.0001$ and $-0.0003$, which are the maximum and minimum values when
$m_{b,c}$ are changed within $1\sigma$ in the analysis.
The input values 
$\bar{m}_c(\bar{m}_c)$ = 1.27($^{+07}_{-11}$)~GeV and
$\bar{m}_b(\bar{m}_b)$ = 4.20($^{+17}_{-07}$)~GeV are taken from
\cite{Amsler:2008zzb}.

\subsection{Lattice spacing}
The uncertainty of the lattice spacing is the largest source of the
systematic error.

Our main result is quoted with the Sommer scale $r_0$ = 0.49~fm as an
input, with which we obtain $a^{-1}$ = 1.83(1)~GeV.
This quantity is convenient because its numerical calculation is
very precise and also because its sea quark mass dependence is mild.
On the other hand, $r_0$ does not have a direct relation to any
physical observables and one has to resort to some model to fix the
central value.
For this reason, one prefers other physical quantities to set the
scale. 

One possible candidate is the pion decay constant $f_\pi$, which is
also precisely measurable on the lattice at unphysical values of sea 
quark masses.
The problem for this quantity is that it may have rather non-trivial
sea quark mass dependence as predicted by Chiral Perturbation Theory
(ChPT). 
Using the next-to-next-to-leading order formula in ChPT, we obtain
$a^{-1}$ = 1.97(4)~GeV \cite{Noaki:2009sk}.
In this analysis we observed a rather large dependence on the sea
quark mass and more importantly a strong curvature that bends the
extrapolation to the lower value of $af_\pi$, thus the higher value of
$a^{-1}$. 
We therefore need more careful analysis on possible systematic errors
in this determination.

One of the most popular quantities to set the scale in recent lattice
calculations is the $\Omega$ baryon mass.
Since the $\Omega$ baryon is made of three strange quarks, the
dependence on up and down quark masses only comes from quark loop
effect, which is expected to be small.
A possible problem is that its determination has to be combined with
the strange quark mass determination, which is non-trivial.
In addition, the finite volume effect could be more important for
baryons. 
Our result is 
$a^{-1}$ = 1.76(8)($^{+5}_{-0}$)~GeV,
with the second error being our estimate of the finite volume effect,
that is set by calculations on a larger volume lattice 
($24^3\times 48$) but at limited values of sea quark mass.

Since each determination has its own advantage and disadvantage, we
decided to take $r_0$ as our central value and others ($f_\pi$ and
$M_\Omega$) to estimate the systematic uncertainties.
The shift of $\alpha_s^{(5)}(M_Z)$ due to the choice of $f_\pi$ and
$M_\Omega$ is $+0.0013$ and $-0.0010$, respectively, which we quote as
the systematic error from the scale setting.

This uncertainty also affects the matching points $m_c$ and $m_b$,
which is included in the above error band.

\subsection{Final result}

\begin{table}
  \begin{center}
    \begin{tabular}{cc}
      \hline\hline
      Sources & Estimated error in $\alpha_s^{(5)}(M_z)$ \\
      \hline
      Uncorrelated fit                             & $\pm$0.0003 \\
      Lattice artifact ($\mathcal{O}(a^2)$ effect) & $+$0.0003 \\
      $\Delta^{V+A}_{\mu\nu}$                        & $\pm$0.0002 \\
      Quark condensate                             & $\pm$0.0001 \\
      $Z_m$                                        & $\pm$0.0001 \\
      Perturbative expansion                       & $\pm$0.0003 \\
      $1/Q^2$ expansion                            & $<$ 0.0001 \\
      $m_{c,b}$                                     & $^{+0.0001}_{-0.0003}$ \\
      Lattice spacing                              & $^{+0.0013}_{-0.0010}$ \\
      \hline
      Total (in quadrature) &                      $^{+0.0014}_{-0.0012}$ \\
      \hline\hline
    \end{tabular}
    \caption{
      Summary of systematic errors in $\alpha_s^{(5)}(M_Z)$.
    }
    \label{tab:systematic_errors}
  \end{center}
\end{table}

Table~\ref{tab:systematic_errors} shows a summary of our estimate of
the systematic errors in our determination of $\alpha_s(M_Z)$. 
We quote final result of the strong coupling constant at the $Z$ boson
mass scale as
\begin{equation} 
  \alpha_s^{(5)}(M_Z)=0.1181(3)(^{+14}_{-12}).
\end{equation}
Here, the first error is statistical error and the second is a sum of
the various systematic errors in quadrature.

This result is consistent with other recent lattice QCD results
0.1174(12) \cite{Allison:2008xk},
0.1183(8) \cite{Davies:2008sw},
0.1192(11) \cite{Maltman:2008bx},
0.1205(8)(5)($^{+\ 0}_{-17}$)\cite{Aoki:2009tf},
and with the world average 0.1184(7) including various high-energy
experiments \cite{Bethke:2009, Amsler:2008zzb} (updated on-line in 2010).

\section{Conclusions}
\label{sec:conclusion}
Determination of the strong coupling constant $\alpha_s(\mu)$ may be
achieved through a perturbative expansion of any physical quantity in
terms of $\alpha_s(\mu)$ at a given scale $\mu$.
Experimental determination typically uses a perturbative amplitude of
quarks at high energy.
Comparison with the lattice QCD calculation provides a highly
non-trivial test of QCD, as lattice uses low-energy hadron spectrum or
matrix elements to set the scale. 

In the lattice calculation, there are variety of choices for the
quantity to be expanded in $\alpha_s(\mu)$.
In order to achieve good enough accuracy, the perturbative expansion
must be known to higher orders or evaluated at very high energies.
The latter may be achieved by calculating the scaling towards the high
energy regime non-perturbatively using the so-called step-scaling
technique (see, {\it e.g.} \cite{DellaMorte:2004bc,Aoki:2009tf}).
The former is numerically less intensive but requires analytic
perturbative calculations beyond one-loop level.

This work demonstrates that the vacuum polarization function can be
used for the precise determination of $\alpha_s$.
The important points are
(i) the perturbative expansion can be done in the continuum theory and
is known to $\mathcal{O}(\alpha_s^3)$,
and
(ii) the non-perturbative lattice calculation with controlled
systematic errors is possible.
The discretization error was a concern as the large $Q^2$ points
have to be calculated, but it turned out to be under control with
currently available lattice setups by careful estimates of systematic
effects. 

The use of the overlap fermion is certainly desirable as the massless
limit of quarks is uniquely defined and the use of the continuum OPE
is justified.
With the lattice fermions that violates chiral symmetry, one expects
dangerous terms such as $m a^{-3}/Q^4$, whose numerical impact has
to be carefully studied.

Extension of this work is straightforward.
Since the largest uncertainty comes from the scale determination, a
consistent determination of the lattice scale with various low-energy
inputs is necessary in order to significantly improve the accuracy.
This requires extensive simulations at larger volumes, smaller quark
masses and smaller lattice spacings.
The discretization effect in VPF considered in this work will also be
significantly reduced by going to finer lattice that will become
available within a few years.

\begin{acknowledgments}
E.S. thanks Y. Kikukawa for correspondence and discussion.
The work of HF was supported by the Global COE program
of Nagoya University called ``QFPU'' from JSPS and MEXT of Japan.
Numerical calculations are performed on IBM System Blue Gene Solution
and Hitachi SR11000 at High Energy Accelerator Research Organization (KEK) 
under a support of its Large Scale Simulation Program (No.~07-16).
This work is supported by the Grant-in-Aid of the Japanese Ministry of Education
(No. 
     18740167, 
     19540286, 
     19740121, 
     20105005, 
     20340047, 
     20105001, 
     20105002, 
     20105003, 
     20025010, 
     21105508, 
     21674002, 
     21684013).  
\end{acknowledgments}


\begin{thebibliography}{99}

\bibitem{Bethke:2009}
  S.~Bethke, 
  Eur. Phys. J. {\bf C} 64, 689 (2009).

\bibitem{Amsler:2008zzb}
  C.~Amsler {\it et al.}  [Particle Data Group],
  Phys.\ Lett.\  B {\bf 667}, 1 (2008).

\bibitem{Davier:2005xq}
  M.~Davier, A.~Hocker and Z.~Zhang,
  Rev.\ Mod.\ Phys.\  {\bf 78}, 1043 (2006);
  [arXiv:hep-ph/0507078].

\bibitem{Shintani:2008ga}
  E.~Shintani {\it et al.}  [JLQCD Collaboration and TWQCD Collaboration],
  Phys.\ Rev.\  D {\bf 79}, 074510 (2009)
  [arXiv:0807.0556 [hep-lat]].

\bibitem{DellaMorte:2004bc}
  M.~Della Morte, R.~Frezzotti, J.~Heitger, J.~Rolf, R.~Sommer and U.~Wolff
                  [ALPHA Collaboration],
  Nucl.\ Phys.\  B {\bf 713}, 378 (2005);
  [arXiv:hep-lat/0411025].

\bibitem{Gockeler:2005rv}
  M.~Gockeler, R.~Horsley, A.~C.~Irving, D.~Pleiter, P.~E.~L.~Rakow, G.~Schierholz and H.~Stuben,
  Phys.\ Rev.\  D {\bf 73}, 014513 (2006).

\bibitem{Aoki:2008tq}
  S.~Aoki {\it et al.}  [JLQCD Collaboration],
  Phys.\ Rev.\  D {\bf 78}, 014508 (2008)
  [arXiv:0803.3197 [hep-lat]].

\bibitem{Chiu:2002}
  T.~W.~Chiu, T.~H.~Hsieh, C.~H.Huang and T.~R.~Huang,
  Phys. Rev. {\bf D} 66, 114502 (2002).

\bibitem{Allison:2008xk}
  I.~Allison {\it et al.},
  Phys. Rev. D {\bf 78}, 054513 (2008).

\bibitem{Kikukawa:1998py}
  Y.~Kikukawa and A.~Yamada,
  Nucl.\ Phys.\  B {\bf 547}, 413 (1999)
  [arXiv:hep-lat/9808026].

\bibitem{Neuberger:1997fp}
  H.~Neuberger,
  Phys.\ Lett.\ B {\bf 417}, 141 (1998)
  [arXiv:hep-lat/9707022].

\bibitem{Neuberger:1998wv}
  H.~Neuberger,
  Phys.\ Lett.\ B {\bf 427}, 353 (1998)
  [arXiv:hep-lat/9801031].

\bibitem{Hashimoto:2008fc}
  S.~Hashimoto,
  arXiv:0811.1257 [hep-lat].

\bibitem{Aoki:2007ka}
  S.~Aoki, H.~Fukaya, S.~Hashimoto and T.~Onogi,
  Phys.\ Rev.\  D {\bf 76}, 054508 (2007)
  [arXiv:0707.0396 [hep-lat]].

\bibitem{Noaki:2009xi}
  J.~Noaki {\it et al.},
  arXiv:0907.2751 [hep-lat].

\bibitem{Surguladze:1990tg}
  L.~R.~Surguladze and M.~A.~Samuel,
  Phys.\ Rev.\ Lett.\  {\bf 66}, 560 (1991)
  [Erratum-ibid.\  {\bf 66}, 2416 (1991)].

\bibitem{Gorishnii:1990vf}
  S.~G.~Gorishnii, A.~L.~Kataev and S.~A.~Larin,
  Phys.\ Lett.\  B {\bf 259}, 144 (1991).

\bibitem{Chetyrkin:1996}
  K. G. Chetyrkin, J. H. K{\" u}hn and M. Steinhauser, 
  Nucl.\ Phys.\ B {\bf 482}, 213 (1996).

\bibitem{Chetyrkin:1993}
  K. G. Chetyrkin, A. Kwiatkowski,
  Z.\ Phys.\ C {\bf 59}, 525 (1993).

\bibitem{Chetyrkin:1985kn}
  K.~G.~Chetyrkin, V.~P.~Spiridonov and S.~G.~Gorishnii,
  Phys.\ Lett.\  B {\bf 160}, 149 (1985).

\bibitem{Martinelli:1996pk}
  G.~Martinelli and C.~T.~Sachrajda,
  Nucl.\ Phys.\  B {\bf 478}, 660 (1996);
  [arXiv:hep-ph/9605336].

\bibitem{Noaki:2008iy}
  J.~Noaki {\it et al.} [JLQCD and TWQCD Collaborations],
  Phys. Rev. Lett. {\bf 101}, 202004 (2008);
  [arXiv:0806.0894 [hep-lat]].

\bibitem{Fukaya:2009fh}
  H.~Fukaya {\it et al.}  [JLQCD collaboration],
  arXiv:0911.5555 [hep-lat].

\bibitem{van Ritbergen:1997va}
  T.~van Ritbergen, J.~A.~M.~Vermaseren and S.~A.~Larin,
  Phys.\ Lett.\  B {\bf 400}, 379 (1997)

\bibitem{Czakon:2004bu}
  M.~Czakon,
  Nucl.\ Phys.\  B {\bf 710} (2005) 485
  [arXiv:hep-ph/0411261].

\bibitem{Gasser:1985}
  J. Gasser and H. Leutwyler, 
  Nucl. Phys. B {\bf 250}, 465 (1985).

\bibitem{Jamin:2002}
  M. Jamin, 
  Phys. Lett. {\bf B} 538, 71 (2002).

\bibitem{Chetyrkin:2006}
  K. G. Chetyrkin, J. H. K\"uhn and C. Sturm,
  Eur. Phys. J. C{\bf 48}, 107 (2006).

\bibitem{Boughezal:2006px}
  R.~Boughezal, M.~Czakon and T.~Schutzmeier,
  Phys.\ Rev.\  D {\bf 74} (2006) 074006
  [arXiv:hep-ph/0605023].

\bibitem{Noaki:2009sk}
  J.~Noaki {\it et al.}  [JLQCD and TWQCD Collaborations],
  PoS {\bf LAT2009}, 096 (2009)
  [arXiv:0910.5532 [hep-lat]].

\bibitem{Davies:2008sw}
  C.~T.~H.~Davies, K.~Hornbostel, I.~D.~Kendall, G.~P.~Lepage, C.~McNeile, J.~Shigemitsu and H.~Trottier
                  [HPQCD Collaboration],
  Phys.\ Rev.\  D {\bf 78}, 114507 (2008)
  [arXiv:0807.1687 [hep-lat]].

\bibitem{Maltman:2008bx}
  K.~Maltman, D.~Leinweber, P.~Moran and A.~Sternbeck,
  Phys.\ Rev.\  D {\bf 78}, 114504 (2008)
  [arXiv:0807.2020 [hep-lat]].

\bibitem{Aoki:2009tf}
  S.~Aoki {\it et al.}  [PACS-CS Collaboration],
  JHEP {\bf 0910}, 053 (2009)
  [arXiv:0906.3906 [hep-lat]].

\end{thebibliography}
\end{document}